\documentclass[12pt]{article}
\usepackage[dvips]{graphics}
\usepackage{graphicx}
\title{Nonlocal black-hole thermodynamics and massive remnants}
\author{Hrvoje Nikoli\'c \\
Theoretical Physics Division, Rudjer Bo\v{s}kovi\'{c} Institute, \\
P.O.B. 180, HR-10002 Zagreb, Croatia \\
{\normalsize hrvoje@thphys.irb.hr} \\
\makebox[1in]{} \\
}
\date{\today}
\begin{document}
\maketitle
\begin{abstract}
To alleviate the black-hole (BH) information problem,
we study a holographic-principle-inspired nonlocal model
of Hawking radiation in which radiated particles created at different 
times all have the same temperature
corresponding to the instantaneous BH mass. Consequently, 
the black hole loses mass not only by continuously radiating 
new particles, but also by continuously warming previously radiated 
particles. The conservation of energy implies that the radiation 
stops when the mass of the black hole reaches the half of the initial 
BH mass, leaving a massive BH remnant 
with a mass much above the Planck scale.
\end{abstract}
\vspace*{0.5cm}
\vspace*{0.9cm}

\noindent
There are two (not necessarily independent) problems related to the 
information contained in black holes. The first one is to explain 
why the information contained in the black hole is proportional 
to its surface, rather than to its volume. The second one is to reconcile 
the process of Hawking radiation with the principle of unitary 
evolution, according to which information cannot be destroyed.
 
A paradigm introduced as a theoretical
framework for dealing with the first problem is 
the holographic principle \cite{hooft,sussk,bouso}, 
according to which the information 
that can be stored inside the black hole is determined by its boundary 
- the black-hole (BH) horizon.
One immediate consequence of the holographic principle is 
{\em nonlocality} - in some way, the degrees of freedom inside the black hole 
should know about the boundary. To explain the holographic principle, 
one must go beyond local quantum field theory. There are indications 
that string theory possesses certain nonlocal features 
(see e.g. \cite{mald,seib,hor,nikolstring}), but the fact is 
that there is not yet a general well-understood theory of 
holography. Instead, the holographic principle often serves merely as a 
guiding principle in construction of physical models. 

A possible solution of the second problem is a BH remnant 
scenario, according to which the process of Hawking radiation 
stops before the black hole evaporates completely, so that 
all the BH information can be contained in the BH remnant.
A problem with the BH remnant scenario is to find a physical mechanism 
that stops the Hawking radiation. If this mechanism is an effect of 
quantum gravity, one generically expects that the significant  
deviation from semiclassical gravity occurs at the Planck scale, 
so that the mass and radius of the remnant are of the order of the 
Planck mass and Planck distance, respectively. Such a light remnant that 
can contain a huge amount of information is problematic 
\cite{harv,gid,strom} because
light objects that can exist in a huge number of different states 
are expected to be often produced in various physical processes, 
which is not seen in nature. Thus, a {\em massive remnant}
\cite{gidrem} (i.e., a remnant with a mass much larger than 
the Planck mass) seems to be a more attractive possibility, but the problem 
is to find a physical mechanism that makes remnants so massive. 
One possibility is that the light remnant cannot exist without 
the Hawking radiation entangled with the BH interior \cite{nikolbh}, 
so that the total system - the black hole together 
with its Hawking radiation - is not light at all.
However, it would be more appealing if the BH remnant itself
would remain massive.  
  
The purpose of the present work is to suggest a possible nonlocal 
physical mechanism that could stop the process of 
Hawking radiation much before  
reaching the Planck scale, thus leaving a massive BH remnant that can store 
the BH information. The mechanism we suggest
is inspired by the holographic principle and partially 
by the suggestion \cite{gidnl} that there should 
exist a nonlocal connection between the degrees of freedom in the 
BH interior and those outside of the black hole.
The holographic principle suggests that the degrees of freedom 
in the BH interior are determined by its boundary - the BH horizon.
On the other hand, if there is also a nonlocal relation between
some exterior degrees of freedom with those in the BH interior, 
then these exterior degrees of freedom could also be determined 
by the same BH boundary. However, it does not seem reasonable to 
expect that {\em all} exterior degrees of freedom are determined 
by a single BH boundary. (There may be a lot of black holes in the
universe, so one particular black hole cannot play a preferred 
role for all exterior degrees of freedom.) Instead, 
neglecting the interaction of the Hawking radiation with other 
exterior degrees of freedom, we assume 
that only the Hawking radiation radiated from this particular 
black hole remains in a nonlocal contact with its boundary. 
The horizon relevant to the process of Hawking radiation 
is the apparent horizon \cite{viss1,viss2}, which, in the case of Hawking 
radiation, is a time-dependent object. The time-dependent 
temperature $T(t)$ associated with the time-dependent 
apparent horizon is determined by the time-dependent BH mass $M(t)$,
through the relation \cite{hawk1,bd}
\begin{equation}\label{temp}
T=\frac{1}{8\pi M},
\end{equation}
where we use units $\hbar=c=G_{\rm N}=k_{\rm B}=1$.
In the standard semiclassical analysis of the process of 
BH radiation,  when particles
are radiated away from the black hole, then the temperature of
these particles does not change with subsequent BH evolution.
Instead, if these particles do not interact with other exterior degrees
of freedom, their temperature remains the same, despite the fact that
later BH temperature may change. Indeed, this is a consequence of 
locality, according to which radiated particles cannot know 
about possible later changes of the BH temperature. 
However, if radiated particles remain in a nonlocal contact 
with its source - the evolving apparent horizon - 
then it seems reasonable to assume that this nonlocal contact 
could manifest as a nonlocal thermodynamic system
in which {\em all} radiated             
particles have the {\em same} temperature (\ref{temp}) given by the
instantaneous BH mass $M(t)$.
Thus, in addition to the standard backreaction of Hawking radiation 
on the black hole (owing to which the BH mass
decreases, such that the total
energy measured by a distant observer is conserved), 
there is an additional nonlocal backreaction of the horizon on
the previously radiatied particles, 
owing to which the temperature of these radiated particles
increases. 

Now let us see how such a nonlocal backreaction leads to massive 
BH remnants.
The energy of all radiated particles at the time $t$ is 
\begin{equation}\label{e1}
E=N\epsilon ,
\end{equation}
where $N$ is the total number of radiated particles at the time $t$
and $\epsilon$ is the
average energy per particle. According to our assumption, 
all radiated particles have the same 
temperature $T(t)$ at a given time $t$, 
so, for massless particles in a thermal equilibrium,
\begin{equation}\label{e2}
\epsilon=bT,
\end{equation}
where $b\sim 1$ is a dimensionless parameter that depends on the
spin of the particles.
Since the energy must be conserved, at each time $t$ there must be 
\begin{equation}\label{cons}
M+N\epsilon=M_0,
\end{equation}
where $M_0$ is the initial BH mass corresponding to $N=0$.
Inserting (\ref{e2}) and (\ref{temp}) into (\ref{cons}), we obtain 
\begin{equation}\label{qe}
N=\frac{(M_0-M)M}{b'} ,
\end{equation}
where $b'\equiv b/8\pi$.
Eq.~(\ref{qe}) can also be viewed as a quadratic equation for $M$,
with the general solution  
\begin{equation}\label{qsol}
M=\frac{M_0}{2} \pm \sqrt{ \left( \frac{M_0}{2} \right)^2 -b'N } .
\end{equation}
The upper sign is the correct one consistent with the 
requirement that $M=M_0$ when $N=0$. During the evolution
in time, $N$ increases while $M$ decreases, until the number 
of radiated particles attains the critical value
\begin{equation}\label{Nc}
N_{\rm crit}=\frac{M_0^2}{4b'}, 
\end{equation}
which corresponds to the vanishing 
square root in (\ref{qsol}). At this moment, the mass 
of the black hole attains the critical value
\begin{equation}\label{Mc}
M_{\rm crit}=\frac{M_0}{2} .
\end{equation}
Can the mass further decrease after reaching this critical value?
Mathematically, it would be possible only by taking the lower 
sign in (\ref{qsol}) at times after 
the mass has reached the critical value. However,
in this case, after this critical moment of time, 
$N$ should start to decrease.
This would correspond to an inverted process of Hawking radiation, 
in which the outgoing Hawking particles suddenly reverse their direction 
of motion, thus becoming ingoing particles that eventually become 
destroyed when they approach the horizon. 
A sudden inversion of the direction of motion does not seem to be 
physical, so we conclude that
such a further decrease of mass is unphysical. In other words, 
the critical mass (\ref{Mc}) is the {\em smallest possible mass}
of the black hole. The final state of BH radiation is 
a massive remnant with the mass
equal to the half of the initial BH mass. 
Clearly, such a massive remnant solves the problem of 
unitary evolution without leading to overproduction 
of light objects that can store a huge amount of information.
In addition, such massive BH remnants could be responsible
for the existence of dark matter in the universe. 

Now let us study the time evolution of such a radiating black hole.   
The change of the radiation energy $E$ is equal
to the negative change of the BH mass $M$, i.e.
$dE=-dM$. Therefore, (\ref{e1}) implies
\begin{equation}\label{e3}
-dM=\epsilon dN +N d\epsilon.
\end{equation}
Note that the second term on the right-hand side of (\ref{e3}) is 
absent in the standard approach, because, in the
standard approach, one assumes that 
the temperature of the fixed number of 
radiated particles does not change, so that $d\epsilon =bdT=0$ in (\ref{e3}).
In our approach, from (\ref{e2}) and (\ref{temp}) we find 
\begin{equation}\label{Ndif}
N d\epsilon =\frac{-b'N}{M^2}dM .
\end{equation}   
The first term on the right-hand side of (\ref{e3}) 
can be calculated in the same 
way as in the standard scenario. Assuming that the black hole 
radiates as a perfect black body, we use the Stefan-Boltzmann law
\begin{equation}\label{SB}
\epsilon dN=\sigma A T^4 dt ,
\end{equation}
where $\sigma=\pi^2/60$ is the Stefan-Boltzmann constant 
and $A=4\pi R^2=16\pi M^2$ is the BH surface. Thus, 
(\ref{SB}) with (\ref{temp}) can be written as
\begin{equation}\label{SB1}
\epsilon dN=\frac{\sigma'}{M^2}dt ,
\end{equation}
where $\sigma'\equiv\sigma/256\pi^3$. Eq.~(\ref{SB1}) with (\ref{e2}) 
and (\ref{temp}) can be written as 
the differential equation
\begin{equation}\label{Nt}
\frac{dN}{dt}=\frac{\sigma'}{b'}\frac{1}{M} .
\end{equation} 
Similarly, by inserting (\ref{Ndif}) and (\ref{SB1}) into (\ref{e3}), 
we obtain another differential equation
\begin{equation}\label{Mt}
\frac{dM}{dt}=\frac{-\sigma'}{M^2-b'N}.
\end{equation}
Eqs.~(\ref{Nt}) and (\ref{Mt}) represent a system of two 
coupled equations that determine the functions $N(t)$ and 
$M(t)$, with the initial conditions 
$N(0)=0$ and $M(0)=M_0$. From (\ref{Nc}), (\ref{Mc}) and 
(\ref{Mt}), we see that $dM/dt$ diverges at the critical values 
of $N$ and $M$. For a numerical analysis of the system 
of equations (\ref{Nt}) and (\ref{Mt}), it is convenient 
to introduce the rescaled variables
\begin{equation}\label{scale}
\tau=\frac{\sigma'}{M_0^3}t, \;\;\;
m=\frac{M}{M_0}, \;\;\;  
n=\frac{b'}{M_0^2}N .
\end{equation}
Eqs.~(\ref{Nt}) and (\ref{Mt}) then become
\begin{equation}\label{nmt}
\frac{dn}{d\tau}=\frac{1}{m} ,\;\;\;\;
\frac{dm}{d\tau}=\frac{-1}{m^2-n} ,
\end{equation}
with the initial conditions $n(0)=0$, $m(0)=1$, 
which does not involve numerical constants that are not 
of the order of unity. 
Eqs.~(\ref{Nt}) and (\ref{Mt}) can also be solved analytically.
From (\ref{Nt}) and (\ref{Mt}) one obtains the differential equation 
$dN/dM=(N-M^2/b')/M$, the solution of which is given by (\ref{qe}).
Thus, by inserting (\ref{qe}) into (\ref{Mt}), 
we obtain a decoupled differential equation
\begin{equation}\label{Mtd}
\frac{dM}{dt}=\frac{-\sigma'}{2M^2-M_0 M} .
\end{equation} 
This is easily integrated to give
\begin{equation}\label{Mtsol}
\frac{2M^3}{3}-\frac{M_0M}{2}-\frac{M_0^3}{6}=-\sigma't ,
\end{equation}
where the requirement $M(t=0)=M_0$ is incorporated.
This, together with (\ref{Mc}), determines the time needed 
for the black hole to approach the critical mass:
\begin{equation}\label{tc}
t_{\rm crit}=\frac{5}{24}\frac{M_0^3}{\sigma'} .
\end{equation}

Let us compare the results above with those in the standard 
semiclassical scenario of BH radiation. 
In this case, there is no second term on the right hand side of 
(\ref{e3}), so (\ref{Mt}) is replaced by a simpler equation  
$dM/dt=-\sigma'/M^2$. Thus, instead of (\ref{Mtsol}), we obtain 
a simpler solution 
\begin{equation}\label{Mts}
M=\sqrt[3]{M_0^3-3\sigma't}.
\end{equation} 
The critical point at which $dM/dt$ diverges corresponds to 
$M_{\rm crit}=0$ and $t_{\rm crit}=(1/3)M_0^3/\sigma'$. 
The number of radiated particles satisfies (\ref{Nt}) which 
is easily integrated to give
\begin{equation}\label{Nts}
N=\frac{M_0^2-(M_0^3-3\sigma't)^{2/3}}{2b'} ,
\end{equation}
leading to $N_{\rm crit}=M_0^2/2b'$. Comparing it with 
(\ref{Nc}), we see that the number of particles produced 
in the massive remnant scenario is equal to the half of that 
in the standard scenario.
Combining (\ref{Mts}) with (\ref{Nts}) one obtains
\begin{equation}
M=\sqrt{ M_0^2 -2b'N } ,
\end{equation}   
which is to be compared with (\ref{qsol}).
The time evolutions of the BH mass in the two scenarios
(Eqs.~(\ref{Mtsol}) and (\ref{Mts})) are compared in Fig.~1.
\begin{figure}[h]
\includegraphics{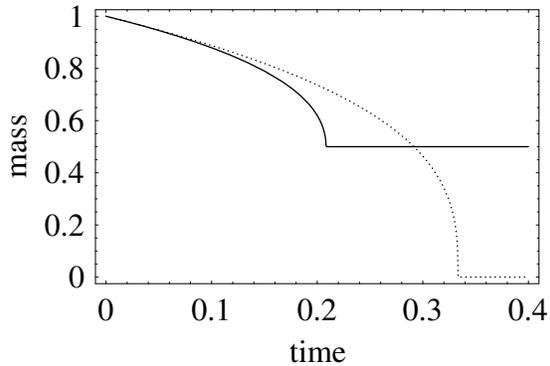}
\caption{\label{fig1}
The BH mass $M/M_0$ as a function of the rescaled time $\tau$.
The solid curve represents the massive remnant scenario, while the
dotted curve represents the standard scenario.}
\end{figure}

As we have seen, the assumption that, at each time, all radiated 
particles have the same temperature given by the instantaneous 
BH mass leads to massive BH remnants. Such remnants offer 
a solution to the problem of unitarity of BH evolution.
The main problem with such a scenario is to find an 
independent justification of such an assumption 
that clearly contradicts the principle of locality.
In particular, this assumption requires a preferred notion 
of the time coordinate, but note that this preferred
time coordinate is the same preferred time coordinate as the one 
needed to define particles in the standard description 
of BH radiation \cite{bd}, which, in an ideal case, 
can be identified with the Killing time.
(Note also that the definition of particles in 
quantum field theory is intrinsically {\em nonlocal} \cite{bd},
even when formulated in terms of local particle currents \cite{nikcur}.)  
We have argued that such a nonlocal thermodynamic behaviour
could be related to the holographic principle, but a 
clear derivation of this relation is missing. Nevertheless,
the fact that such a simple assumption leads to such 
a simple (and surprising!) solution to the problem of BH unitarity
leads us to suspect that this assumption could be on the right track.
Hopefully, the future research could reveal a more 
compelling explanation of our assumption.

\section*{Acknowledgements}

This work was supported by the Ministry of Science and Technology of the
Republic of Croatia.

\end{document}